\documentclass[journal,10pt]{IEEEtran}

\usepackage[utf8]{inputenc}
\usepackage[T1]{fontenc}
\usepackage{graphicx}
\usepackage{textcomp}
\usepackage{xcolor}
\usepackage{booktabs}
\usepackage{multirow}
\usepackage{array}
\usepackage{tabularx}
\usepackage{url}
\usepackage{hyperref}
\usepackage{cite}
\usepackage{enumitem}
\usepackage{tcolorbox}
\usepackage{float}

\tcbuselibrary{skins,breakable}

\hypersetup{
  colorlinks=true,
  linkcolor=black,
  citecolor=black,
  urlcolor=blue
}

\newcolumntype{L}[1]{>{\raggedright\arraybackslash}p{#1}}

\newtcolorbox{keypoint}[1][]{
  colback=blue!5!white,
  colframe=blue!40!black,
  fonttitle=\bfseries,
  title={Key Point},
  #1
}

\newtcolorbox{actionbox}[1][]{
  colback=green!5!white,
  colframe=green!40!black,
  fonttitle=\bfseries,
  title={Action Required},
  #1
}

\newtcolorbox{warningbox}[1][]{
  colback=red!5!white,
  colframe=red!40!black,
  fonttitle=\bfseries,
  title={Warning},
  #1
}

\begin{document}

\title{The UK Cyber Security and Resilience Bill:\\A Practitioner's Guide to Legislative Reform, Compliance,\\and Organisational Readiness}

\author{\IEEEauthorblockN{Jonathan Shelby}\\
\IEEEauthorblockA{Department of Computer Science\\
University of Oxford\\
Oxford, United Kingdom\\
jonathan.shelby@cs.ox.ac.uk}}

\maketitle

\begin{abstract}
The Cyber Security and Resilience (Network and Information Systems) Bill, introduced to Parliament in November 2025, represents the most significant reform of UK cyber security legislation in nearly a decade. This paper provides a comprehensive practitioner-oriented analysis of the Bill's provisions, their practical implications, and the steps organisations must take to achieve compliance. It examines the expanded regulatory scope covering managed service providers, data centres, and designated critical suppliers; the enhanced 24/72-hour incident reporting regime; the strengthened enforcement architecture including penalties of up to \pounds17 million or 4\% of worldwide turnover; and the Secretary of State's new executive powers. The paper compares the Bill with the EU's NIS2 Directive and DORA, proposing a practical dual-compliance framework for financial services firms. It explains how Zero Trust Architecture principles can serve as a foundation for meeting the Bill's requirements, and how the NCSC's Cyber Assessment Framework v4.0 provides the assurance pathway. Four detailed appendices provide entity-specific compliance roadmaps, worked case studies mapping real UK incidents to Bill provisions, sector-specific action plans for financial services, energy, health, and MSPs, and a complete gap analysis and self-assessment tool mapped to CAF v4.0 and the Bill's requirements.
\end{abstract}

\begin{IEEEkeywords}
Cyber Security and Resilience Bill, NIS Regulations, NIS2, DORA, Critical National Infrastructure, Zero Trust, Cyber Assessment Framework, Managed Service Providers, Supply Chain Security, Compliance, Governance
\end{IEEEkeywords}

\section{Introduction}
\label{sec:introduction}

The United Kingdom faces a cyber threat of unprecedented scale and sophistication. In its 2025 Annual Review, the National Cyber Security Centre (NCSC) reported a record 204 nationally significant cyber incidents in the year to September 2025---more than double the 89 recorded in the previous period \cite{ncsc2025annual}. Ransomware and supply chain attacks have disrupted the NHS, local authorities, utilities, and defence systems, with the estimated annual cost to the UK economy approaching \pounds15 billion \cite{traverssmith2025bill}.

\begin{keypoint}
The NCSC recorded 204 nationally significant cyber incidents in 2024--25, averaging four major attacks per week. The UK economy loses an estimated \pounds15 billion annually to cyber attacks.
\end{keypoint}

Against this backdrop, the Cyber Security and Resilience (Network and Information Systems) Bill was introduced to the House of Commons on 12 November 2025 \cite{ukparliament2025bill}. The Bill amends and substantially extends the Network and Information Systems Regulations 2018, which have governed cyber security obligations for critical infrastructure since the UK's transposition of the original EU NIS Directive.

This paper is written for CISOs, heads of information security, compliance officers, board members, and risk professionals across the UK's critical infrastructure, digital services, and financial services sectors. It provides both the strategic context necessary to understand the Bill's significance and the practical tools required to prepare for compliance. The paper is structured as follows:

\begin{itemize}[itemsep=1pt]
\item Sections~\ref{sec:context}--\ref{sec:provisions}: What the Bill says and why it matters
\item Section~\ref{sec:international}: How it compares with EU NIS2 and DORA
\item Sections~\ref{sec:zta}--\ref{sec:caf}: The technical and assurance foundations for compliance
\item Section~\ref{sec:supplychain}: Supply chain security and the new Designated Critical Supplier mechanism
\item Section~\ref{sec:enforcement}: The enforcement architecture and its implications
\item Appendices A--D: Compliance roadmaps, case studies, sector guidance, and self-assessment tools
\end{itemize}

\section{Legislative Context: Why This Bill, Why Now}
\label{sec:context}

\subsection{The Limitations of the Existing Regime}

The NIS Regulations 2018 established statutory cyber security obligations for operators of essential services (OES) across five sectors---energy, transport, health, drinking water, and digital infrastructure---and for relevant digital service providers (RDSPs) including cloud computing services, online marketplaces, and search engines \cite{nisregs2018}. Each sector was assigned a competent authority responsible for oversight.

The regulations represented an important first step, but three critical limitations emerged \cite{hoclib2026briefing}:

\textit{Insufficient scope.} Key service types---most notably managed service providers (MSPs) that underpin the IT operations of critical organisations---fell outside the regulatory perimeter. An MSP could be compromised, disrupting an essential service, without facing any NIS-specific regulatory consequence.

\textit{Inadequate reporting.} Only a fraction of significant incidents were reported to regulators, leaving the Government unable to develop an accurate picture of the national threat landscape.

\textit{Legislative rigidity.} Because the NIS Regulations were made under the European Communities Act 1972 (since repealed), the Government lacked delegated powers to update them without primary legislation.

\subsection{The Incidents That Forced the Issue}

Several high-profile attacks underscored the urgency of reform. The 2023 ransomware attack on Advanced Computer Software disrupted NHS 111 services across multiple trusts \cite{cliffordchance2025}. The 2023 attack on Capita, a major outsourcer to local government, the NHS, and the Ministry of Defence, demonstrated how a single provider's compromise could cascade across multiple sectors. In 2025, attacks on Marks \& Spencer and Jaguar Land Rover---the latter estimated to have cost the UK economy \pounds1.9 billion \cite{traverssmith2025bill}---showed that nationally significant disruption was not confined to the sectors currently regulated under NIS.

\begin{warningbox}
The Advanced/NHS, Capita, M\&S, and JLR incidents collectively demonstrate that the existing NIS Regulations were insufficient in both scope and enforcement. See Appendix~\ref{app:casestudies} for detailed case study analysis.
\end{warningbox}

The NCSC characterised the situation as a ``widening gap between the increasingly complex cyber threats and the UK's defensive capabilities'' \cite{ncsc2025annual}. The incoming Labour Government prioritised the Bill in its July 2024 King's Speech, and the April 2025 Policy Statement confirmed its measures \cite{dsit2025policy}.

\section{Key Provisions of the Bill}
\label{sec:provisions}

\subsection{Expanded Scope: Who Is Now Regulated?}

The Bill's most significant structural change is the expansion of regulated entities. Three new categories are introduced \cite{ukparliament2025bill, dsit2025factsheets}:

\textbf{Relevant Managed Service Providers (RMSPs):} Medium or large businesses (50+ employees or $>$\texteuro10m turnover) providing ongoing management of IT systems---including support, maintenance, monitoring, and active administration---to separate legal entities. Overseen by the ICO.

\textbf{Data Centres:} Brought into scope following their designation as critical national infrastructure, recognising their essential role in the digital economy.

\textbf{Designated Critical Suppliers (DCSs):} Third-party suppliers whose goods or services are essential to the delivery of an OES's essential service, where a cyber incident affecting the supplier could significantly impact that service. Designated by the relevant competent authority.

\begin{keypoint}
The Bill also brings large load controllers---organisations controlling the energy use of smart appliances, batteries, and electric vehicles---into scope, reflecting the growing convergence of IT and operational technology in critical infrastructure.
\end{keypoint}

\subsection{Enhanced Incident Reporting}

The Bill replaces the existing flexible 72-hour reporting window with a mandatory two-stage process \cite{ukparliament2025bill}:

\begin{itemize}[itemsep=1pt]
\item \textbf{Within 24 hours:} An initial notification to the relevant competent authority, serving as an early warning.
\item \textbf{Within 72 hours:} A comprehensive incident report providing full details.
\item \textbf{Concurrent NCSC notification:} Both the regulator and the NCSC receive the same information simultaneously.
\item \textbf{Customer notification:} Digital service providers and data centre operators must alert affected customers.
\end{itemize}

Critically, the definition of ``reportable incident'' now includes events that are \textit{capable of having} a significant impact, not only those that have already caused one. This captures ransomware infections (where systems are impaired pending payment) and pre-positioning events (where attackers establish covert access for future exploitation) \cite{ukparliament2025bill}.

\subsection{Strengthened Enforcement}

The Bill introduces a simplified two-band penalty structure replacing the existing four-tier model \cite{dsit2025enforcement}:

\begin{itemize}[itemsep=1pt]
\item \textbf{Serious breaches:} Up to \pounds17 million or 4\% of worldwide turnover (whichever is higher)
\item \textbf{Less serious breaches:} Up to \pounds10 million or 2\% of worldwide turnover
\item \textbf{Continuing non-compliance:} Daily fines of \pounds100,000 for failure to act on regulatory directions \cite{kyle2025statement}
\end{itemize}

Regulators also gain enhanced investigatory powers, including the ability to conduct proactive vulnerability assessments, and a new cost recovery mechanism to fund their oversight activities from fees charged to regulated entities \cite{ukparliament2025bill}.

\subsection{Secretary of State's Executive Powers}

The Bill confers substantial new powers on the Secretary of State \cite{ukparliament2025bill}:

\begin{itemize}[itemsep=1pt]
\item \textbf{Statement of Strategic Priorities:} Setting the direction for UK cyber resilience, to which all regulators must have regard.
\item \textbf{Statutory Codes of Practice:} Specifying detailed compliance expectations, subject to consultation and parliamentary scrutiny.
\item \textbf{Scope expansion:} Through secondary legislation, designating new essential activities and regulated persons without further primary legislation.
\item \textbf{Emergency directions:} Directing regulators and regulated entities to take specific steps where there is a national security threat, overriding other regulatory obligations.
\end{itemize}

\begin{keypoint}
The emergency direction power mirrors mechanisms in the Telecommunications (Security) Act 2021 \cite{telecomsecact2021} and gives the Government agility to respond to acute threats without waiting for legislative change.
\end{keypoint}

\section{International Context: How the Bill Compares}
\label{sec:international}

\subsection{CS\&R Bill vs EU NIS2 Directive}

The Bill draws on lessons from the EU's NIS2 Directive \cite{nis2directive} whilst maintaining a distinctly UK approach. Table~\ref{tab:comparison} summarises the key differences.

\begin{table*}[!t]
\centering
\caption{Comparative Analysis: UK CS\&R Bill vs EU NIS2 Directive}
\label{tab:comparison}
\begin{tabular}{L{2.8cm}L{5.5cm}L{5.5cm}}
\toprule
\textbf{Dimension} & \textbf{UK CS\&R Bill} & \textbf{EU NIS2 Directive} \\
\midrule
Scope & MSPs, data centres, DCSs added to existing OES/RDSP framework & 17 sectors; essential and important entity categories; approx. 160,000 entities \\
\addlinespace
Incident reporting & 24h initial / 72h full; concurrent NCSC notification & 24h early warning (essential); 72h full report; one-month final report \\
\addlinespace
Penalties & \pounds17m / 4\% turnover (serious); \pounds100k/day for directions & \texteuro10m / 2\% turnover (essential); \texteuro7m / 1.4\% (important) \\
\addlinespace
Board accountability & No explicit director liability; Cyber Governance Code voluntary & Management body directly responsible; mandatory training \\
\addlinespace
Regulatory model & 12 sector-specific regulators & Member State discretion; NIS Cooperation Group \\
\addlinespace
Flexibility & SoS powers to expand scope and issue Codes of Practice & Member States may adopt higher standards \\
\addlinespace
Supply chain & DCS designation mechanism; supply chain duties on OES/RDSP & Supply chain risk management obligations; multi-disciplinary assessment \\
\bottomrule
\end{tabular}
\end{table*}

The most significant divergence is on \textbf{board-level accountability}. NIS2 makes management bodies directly responsible for cyber security and mandates board-level training. The UK Bill contains no equivalent provision, relying instead on the voluntary Cyber Governance Code of Practice and direct Government correspondence to the UK's top 350 businesses \cite{traverssmith2025bill}. For organisations that take governance seriously, this is a gap worth closing voluntarily.

The UK Bill's scope is also \textbf{narrower}: NIS2 covers 17 sectors including manufacturing, food, postal services, and waste management. The UK Bill focuses on existing NIS sectors plus MSPs, data centres, and critical suppliers. The Government's argument is that the Secretary of State's scope expansion powers provide flexibility to extend coverage without primary legislation \cite{hoclib2026briefing}. Critics, pointing to the M\&S and JLR attacks, argue this is too slow \cite{traverssmith2025bill}.

\subsection{Relationship with DORA}

For financial services firms, the EU's Digital Operational Resilience Act (DORA) creates an additional compliance dimension \cite{dora2022}. DORA applies to UK firms providing services into the EU and imposes prescriptive requirements for ICT risk management, resilience testing (including threat-led penetration testing), third-party oversight, and incident reporting.

\begin{actionbox}
UK financial services firms with EU operations face dual compliance obligations under both the CS\&R Bill and DORA. Section~\ref{sec:doraframework} provides a practical framework for managing this overlap. See also Appendix~\ref{app:sectoral} for the financial services action plan.
\end{actionbox}

\subsection{A Practical Dual-Compliance Framework}
\label{sec:doraframework}

Table~\ref{tab:doracsrmap} maps the principal obligations across both regimes.

\begin{table*}[!t]
\centering
\caption{Obligation Mapping: CS\&R Bill vs DORA for UK Financial Services}
\label{tab:doracsrmap}
\begin{tabular}{L{2.4cm}L{3.8cm}L{3.8cm}L{3.4cm}}
\toprule
\textbf{Domain} & \textbf{CS\&R Bill} & \textbf{DORA} & \textbf{Practical Approach} \\
\midrule
Risk management & Outcomes-based via CAF; Codes of Practice forthcoming & Prescriptive ICT risk framework; specific policies mandated & Adopt DORA standard as baseline; CS\&R compliance follows \\
\addlinespace
Incident reporting & 24h initial / 72h full to CA + NCSC & Major ICT incident reporting to ESA & Build unified triage function serving both timelines \\
\addlinespace
Third-party oversight & DCS designation; supply chain duties & TPSP register; concentration risk; ESA oversight of critical TPSPs & Combine into single vendor risk programme \\
\addlinespace
Resilience testing & Not mandated (expected in Codes) & TLPT for significant entities; annual basic testing & Adopt TLPT as baseline across all operations \\
\addlinespace
Governance & No explicit board liability & Management body directly responsible & Adopt DORA governance standard UK-wide \\
\addlinespace
Penalties & \pounds17m / 4\% turnover & Per Member State transposition & Budget for cumulative exposure from both regimes \\
\bottomrule
\end{tabular}
\end{table*}

The guiding principle is straightforward: \textbf{where obligations overlap, adopt the higher standard as the baseline across the entire organisation}. This avoids maintaining parallel compliance programmes and ensures that satisfying DORA (which is generally more prescriptive) automatically achieves CS\&R compliance across the overlapping domains.

In practice, this means four layers of implementation:

\textit{Layer 1 --- Unified Controls:} A single integrated control framework mapped to both CAF v4.0 and DORA's requirements. Test once, evidence once, satisfy both.

\textit{Layer 2 --- Dual Reporting:} Parallel but coordinated incident reporting workflows, with a single triage function that classifies incidents against both regimes' thresholds simultaneously.

\textit{Layer 3 --- Integrated Third-Party Risk:} One vendor risk management programme combining DORA's TPSP register, the CS\&R Bill's DCS mechanism, and the Bank of England's critical third-party regime \cite{boe2021ss2}.

\textit{Layer 4 --- Unified Governance:} Even though the CS\&R Bill doesn't mandate it, adopt DORA's board accountability standard across UK operations. This is the right thing to do, and it positions the organisation well for the likely direction of future UK legislation.

\section{Zero Trust as a Compliance Foundation}
\label{sec:zta}

\subsection{Why Zero Trust Aligns with the Bill}

Zero Trust Architecture (ZTA) operates on the principle of ``never trust, always verify''---enforcing continuous authentication, conditional access, and least-privilege controls across all users, devices, and services, regardless of their location on the network \cite{nist800207}. This approach aligns directly with several of the Bill's core requirements:

\textbf{Continuous verification} supports the Bill's incident detection and reporting obligations. If you are continuously evaluating the trust status of every connection, you are far more likely to detect anomalies---including the pre-positioning events that the Bill's expanded incident definition now captures.

\textbf{Micro-segmentation} limits the blast radius of supply chain compromises. When a managed service provider's credentials are compromised, micro-segmentation ensures the attacker cannot move laterally across the entire network. This directly supports the supply chain risk management that underpins the Bill's DCS mechanism.

\textbf{Least-privilege access} ensures that no user or service has more access than is strictly necessary for their function. This reduces the impact of compromised accounts and aligns with the CAF v4.0 outcomes on identity verification and privileged user management.

\textbf{Assume breach} is the foundational mindset of Zero Trust: design your systems on the assumption that a breach has already occurred, and ensure that your architecture limits the damage. This aligns with the Bill's emphasis on resilience, not merely prevention.

\begin{keypoint}
Zero Trust is not a product you buy. It is an architectural philosophy and a set of design principles that inform how you build, configure, and operate your systems. Adoption is a journey, not a destination, and it can be implemented incrementally.
\end{keypoint}

\subsection{Practical ZTA Adoption for Bill Compliance}

The NIST SP 800-207 framework \cite{nist800207} and the DoD Zero Trust Reference Architecture v2.0 \cite{dodztra2022} provide structured approaches to adoption. For organisations preparing for the CS\&R Bill, the following priorities are recommended:

\textbf{1. Identity as the new perimeter.} Invest in strong identity and access management: multi-factor authentication for all administrative access, conditional access policies that evaluate device posture and location, and privileged access management with just-in-time elevation.

\textbf{2. Micro-segment supply chain access.} Where MSPs or other third parties have access to your systems, segment their access into the minimum necessary scope. Monitor all third-party sessions continuously. This directly reduces the dependency risk captured in the Bill's DCS framework.

\textbf{3. Continuous monitoring and logging.} Implement comprehensive logging with defined retention periods, security information and event management (SIEM), and automated alerting for anomalous behaviour. This provides both the detection capability the Bill expects and the evidence base for regulatory reporting.

\textbf{4. Data classification and encryption.} Classify data by sensitivity, encrypt at rest and in transit, and implement data loss prevention controls. This supports CAF v4.0 Objective B outcomes on data security.

\textbf{5. Resilient architecture.} Design systems with redundancy and failover capability, tested regularly. Assume that any component---including a critical supplier---may fail, and ensure that the overall service continues.

\subsection{Mapping ZTA to CAF v4.0}

Table~\ref{tab:ztacaf} illustrates how Zero Trust principles map to the NCSC Cyber Assessment Framework, providing a practical bridge between architectural decisions and regulatory compliance.

\begin{table*}[!t]
\centering
\caption{Mapping Zero Trust Principles to CAF v4.0 Contributing Outcomes}
\label{tab:ztacaf}
\begin{tabular}{L{3.0cm}L{3.2cm}L{7.0cm}}
\toprule
\textbf{ZTA Principle} & \textbf{CAF v4.0 Outcome} & \textbf{How It Supports Compliance} \\
\midrule
Continuous authentication & B2a: Identity verification \& authentication & Demonstrates that access decisions are based on real-time trust evaluation, not static credentials \\
\addlinespace
Least-privilege access & B2c: Privileged user management & Evidences that access is proportionate and dynamically enforced \\
\addlinespace
Micro-segmentation & B5: Resilient networks \& systems & Limits blast radius; provides evidence of proportionate supply chain controls \\
\addlinespace
Continuous monitoring & C1: Security monitoring & Supports detection capability and provides audit trail for incident reporting \\
\addlinespace
Assume breach / resilience & D1--D4: Response \& recovery & Ensures architecture supports rapid detection, containment, and recovery \\
\addlinespace
Data-centric security & B3: Data security & Demonstrates that data protection is embedded in architecture, not bolted on \\
\bottomrule
\end{tabular}
\end{table*}

\section{The NCSC Cyber Assessment Framework v4.0}
\label{sec:caf}

\subsection{The Assurance Backbone}

The NCSC Cyber Assessment Framework is the principal assurance mechanism for evaluating cyber resilience among UK organisations responsible for essential functions. CAF v4.0, released in August 2025 \cite{ncsc2025cafv4}, provides the structured approach that competent authorities will use to assess whether regulated entities are meeting their obligations under the Bill.

The Framework is built around four objectives and 14 principles, each defined in terms of outcomes rather than prescriptive controls \cite{ncsc2025cafintro}:

\textbf{Objective A: Managing security risk} --- governance, risk management, asset management, and supply chain oversight.

\textbf{Objective B: Protecting against cyber attack} --- identity and access management, data security, system security, and resilient networks.

\textbf{Objective C: Detecting cyber security events} --- security monitoring, anomaly detection, and threat intelligence.

\textbf{Objective D: Minimising the impact of incidents} --- response planning, recovery capability, and lessons learned.

These four objectives decompose into 41 contributing outcomes, each assessed using Indicators of Good Practice (IGPs) that support a determination of ``Achieved'', ``Partially Achieved'', or ``Not Achieved'' \cite{ncsc2025cafv4}.

\subsection{What Changed in v4.0}

CAF v4.0 introduces several enhancements directly relevant to the Bill's trajectory \cite{ncsc2025cafv4}:

\begin{itemize}[itemsep=1pt]
\item Strengthened guidance on \textbf{threat modelling}, reflecting the Bill's emphasis on proportionate, risk-informed security
\item New expectations on \textbf{software security lifecycle management}, addressing supply chain software risks
\item Enhanced \textbf{threat hunting} guidance, supporting the Bill's expanded incident detection expectations
\item New \textbf{AI governance} considerations, anticipating the growing use of AI in both attacks and defences
\end{itemize}

\subsection{CAF Profiles}

A CAF Profile is a curated subset of contributing outcomes at specified achievement levels, set by the relevant competent authority to define proportionate security expectations for a sector \cite{ncsc2025cafintro}. The Basic Profile sets the minimum standard for resilience against common attacks. Enhanced Profiles address more sophisticated threat scenarios. Organisations should engage with their competent authority to understand which profile applies and plan accordingly.

\begin{actionbox}
If your organisation has not yet conducted a CAF assessment, this should be an immediate priority. The gap analysis tool in Appendix~\ref{app:gapanalysis} provides a structured starting point mapped to both CAF v4.0 and the Bill's specific requirements.
\end{actionbox}

\section{Supply Chain Security and the DCS Mechanism}
\label{sec:supplychain}

\subsection{The Scale of the Problem}

Supply chain attacks are growing faster than any other threat vector. Verizon's 2024 Data Breach Investigations Report found that incidents involving software vulnerabilities increased by 180\% year-on-year, with 15\% involving a third-party supplier \cite{verizon2024dbir}. Between 2021 and 2023, supply chain attacks increased by 431\% globally. The ISC2 2025 Supply Chain Risk Survey found 70\% of cyber security professionals reported high concern about supply chain risks \cite{isc22025roundtable}.

\subsection{How the DCS Mechanism Works}

The Bill's Designated Critical Supplier mechanism enables competent authorities to identify and regulate specific suppliers whose compromise could cascade to affect essential services \cite{dsit2025factsheets}. A supplier may be designated if three conditions are met:

\begin{enumerate}[itemsep=1pt]
\item The supplier provides goods or services directly to an OES.
\item The supplier relies on network and information systems for that supply.
\item An incident affecting the supplier's NIS could significantly impact the OES's service delivery.
\end{enumerate}

Once designated, the supplier faces direct security obligations and regulatory oversight from the designating competent authority \cite{traverssmith2025bill}.

\subsection{Assessing Your Supply Chain Exposure}

Organisations should assess their supply chain risk across three dimensions:

\textbf{Business criticality:} How essential is this supplier's service to the delivery of your essential function? If the supplier ceased operations tomorrow, could you continue?

\textbf{Substitutability:} Are there readily available alternative suppliers who could provide an equivalent service within an acceptable timeframe? Sole-source dependencies represent the highest risk.

\textbf{Downstream resilience:} If this supplier were compromised, how effectively could your organisation contain the impact? This is where micro-segmentation and Zero Trust access controls directly reduce risk.

The combination of high criticality, low substitutability, and low resilience identifies the suppliers most likely to be designated as DCSs---and the relationships that most urgently require strengthening.

\begin{keypoint}
The cascade effect is the critical risk: a single compromised MSP can simultaneously disrupt multiple OES across different sectors, as the Capita incident demonstrated. Organisations should map their full dependency chain, not merely their direct suppliers.
\end{keypoint}

\subsection{Practical Mitigation}

Zero Trust micro-segmentation provides the most effective technical mitigation against supply chain cascade risk. By isolating each supplier's access into independent network segments, an organisation ensures that the compromise of one supplier's credentials cannot be leveraged to access other systems or data. The more independent segments you maintain, the exponentially greater your resilience---each additional segment multiplicatively reduces the probability of a complete system compromise through a single supplier.

Contractual controls are equally important: supply agreements should include provisions for incident notification, regulatory cooperation, audit access, security assurance reporting, and exit strategies that enable service continuity in the event of a supplier's regulatory failure.

\section{The Enforcement Architecture}
\label{sec:enforcement}

\subsection{The Twelve-Regulator Model}

The Bill assigns oversight to twelve sector-specific competent authorities \cite{dsit2025factsheets, reedsmith2026implications}: Ofgem (energy), the Department for Transport (transport), the Department of Health and Social Care (health), the Drinking Water Inspectorate (water), Ofcom (digital infrastructure and telecoms), and the ICO (MSPs, RDSPs, and data centres), along with additional regulators for subsectors including the Civil Aviation Authority, the Maritime and Coastguard Agency, and the Health and Safety Executive.

\subsection{The Consistency Challenge}

The multi-regulator model brings sector expertise but risks inconsistent implementation. Each regulator has different institutional cultures, enforcement traditions, levels of cyber security expertise, and resources \cite{hoclib2026briefing}. Specific risks include:

\begin{itemize}[itemsep=1pt]
\item Divergent interpretations of what constitutes a ``significant incident'' across sectors
\item Varying expectations for technical and organisational measures
\item Inconsistent cost recovery fee structures
\item Different enforcement thresholds creating uneven playing fields
\end{itemize}

An MSP serving both healthcare and energy clients could face different compliance expectations from different regulators for the same service. The Statement of Strategic Priorities and statutory Codes of Practice provide alignment mechanisms, but their effectiveness will depend on the quality of inter-regulator coordination.

\subsection{The Case for Board-Level Accountability}

The Bill's omission of explicit director-level accountability is its most significant governance gap relative to NIS2 and to the trajectory of UK regulatory thinking in financial services. The Senior Managers and Certification Regime (SM\&CR) provides a proven precedent: senior managers bear personal responsibility for the areas they oversee, with regulators empowered to act against individuals who fail to take reasonable steps \cite{fca2022opres}.

\begin{warningbox}
Even without statutory board-level liability under the CS\&R Bill, directors remain subject to general corporate law duties of care and diligence. Organisations should not mistake the absence of an SM\&CR-equivalent for an absence of personal accountability.
\end{warningbox}

Our recommendation is clear: regardless of the Bill's current position, organisations should voluntarily designate a senior individual with explicit accountability for NIS compliance. This is the direction of travel, and proactive adoption positions the organisation favourably for likely future amendments.

\section{Conclusions and Recommendations}
\label{sec:conclusions}

The CS\&R Bill represents a substantial and necessary modernisation of the UK's cyber security legislative framework. Its expanded scope, enhanced reporting obligations, strengthened enforcement, and flexible executive powers collectively establish a more robust architecture for protecting critical national infrastructure.

For practitioners, the key messages are:

\begin{enumerate}[label=\textbf{\arabic*.}, itemsep=3pt]
\item \textbf{Act now, not at Royal Assent.} The Bill's direction is clear and its provisions are unlikely to be weakened during parliamentary passage. Organisations that begin compliance preparation now will avoid the rush and achieve higher quality outcomes.

\item \textbf{Use CAF v4.0 as your roadmap.} The Cyber Assessment Framework provides the structured pathway to demonstrating compliance. Conduct a gap analysis (Appendix~\ref{app:gapanalysis}), develop a remediation plan, and establish a regular assessment cycle.

\item \textbf{Adopt Zero Trust principles.} Not as a product procurement exercise, but as an architectural philosophy that directly supports the Bill's requirements for continuous verification, supply chain segmentation, and resilient systems.

\item \textbf{Map your supply chain.} The DCS mechanism will reshape supply chain relationships across UK critical infrastructure. Understand your dependencies, assess their criticality and substitutability, and strengthen the most vulnerable relationships.

\item \textbf{Prepare for the reporting regime.} The 24/72-hour dual-notification timeline is demanding. Test it now with tabletop exercises, before it becomes a statutory obligation.

\item \textbf{Engage your board.} Even without statutory board liability, cyber resilience is a board-level issue. The Bill's penalty structure (\pounds17m / 4\% turnover) ensures that non-compliance is a material risk to any organisation.

\item \textbf{For financial services: build for dual compliance.} Adopt the higher DORA standard as your baseline across overlapping domains. This avoids maintaining parallel compliance programmes and future-proofs against evolving UK requirements.

\item \textbf{Engage your regulators early.} Proactive dialogue with competent authorities enables better understanding of forthcoming Codes of Practice, expected CAF profiles, and the timeline for secondary legislation.
\end{enumerate}

The Bill is expected to progress through Parliament and commence in phases following Royal Assent. For organisations across the UK's critical infrastructure and digital services landscape, the imperative is clear: the organisations that prepare proactively will be those that thrive under the new regime.

\bibliographystyle{IEEEtran_custom}
\bibliography{references}

\clearpage
\appendices

\section{Compliance Readiness Checklists and Implementation Roadmaps}
\label{app:roadmaps}

This appendix provides entity-specific implementation roadmaps aligned with the anticipated commencement timeline. The roadmaps assume Royal Assent in mid-2026, with phased commencement through secondary legislation. Adjust timelines as the parliamentary timetable crystallises.

\subsection{Operators of Essential Services (OES)}

\subsubsection{Phase 1: Immediate (0--6 months post-Royal Assent)}
\begin{enumerate}[label=\textbf{OES-\arabic*.}, leftmargin=*, itemsep=2pt]
\item \textbf{Board briefing and mandate.} Brief the board on enhanced requirements and the \pounds17m / 4\% turnover penalty structure. Secure budget and accountability for a compliance programme.
\item \textbf{Competent authority engagement.} Initiate dialogue with the relevant CA to understand forthcoming Codes of Practice, expected CAF profiles, and secondary legislation timelines.
\item \textbf{Gap analysis against CAF v4.0.} Conduct a comprehensive assessment using the tool in Appendix~\ref{app:gapanalysis}. Prioritise remediation by risk severity.
\item \textbf{Incident reporting process review.} Map existing procedures against the 24/72-hour timeline. Identify bottlenecks preventing timely dual notification to the CA and NCSC.
\item \textbf{Supply chain inventory.} Compile a complete register of third-party suppliers relying on NIS that are material to essential service delivery.
\end{enumerate}

\subsubsection{Phase 2: Medium-Term (6--18 months)}
\begin{enumerate}[label=\textbf{OES-\arabic*.}, leftmargin=*, start=6, itemsep=2pt]
\item \textbf{Supply chain risk assessment.} Assess each supplier's criticality, substitutability, and downstream resilience. Engage with the CA on potential DCS designations.
\item \textbf{Incident response rehearsal.} Conduct at least two tabletop exercises including ransomware and pre-positioning scenarios. Test the dual-notification workflow.
\item \textbf{Technical controls uplift.} Implement ZTA principles: continuous authentication, micro-segmentation of supply chain access, and continuous monitoring.
\item \textbf{Code of Practice alignment.} Implement requirements from the first statutory Codes of Practice. Map to existing controls and close residual gaps.
\item \textbf{Cost recovery planning.} Engage with the CA regarding the anticipated fee structure. Budget for regulatory fees.
\end{enumerate}

\subsubsection{Phase 3: Steady State (18+ months)}
\begin{enumerate}[label=\textbf{OES-\arabic*.}, leftmargin=*, start=11, itemsep=2pt]
\item \textbf{Ongoing CAF cycle.} Establish at minimum annual self-assessment, supplemented by periodic independent assessment.
\item \textbf{Continuous improvement.} Embed improvement driven by incident lessons, threat intelligence, and CAF findings.
\item \textbf{Supply chain monitoring.} Ongoing monitoring of DCSs including contractual reviews, security assessments, and incident coordination.
\end{enumerate}

\subsection{Relevant Managed Service Providers (RMSPs)}

\subsubsection{Phase 1: Immediate (0--6 months)}
\begin{enumerate}[label=\textbf{MSP-\arabic*.}, leftmargin=*, itemsep=2pt]
\item \textbf{Scope determination.} Confirm whether the RMSP definition applies: $\geq$50 employees or $>$\texteuro10m turnover; ongoing IT management to separate entities using NIS. Seek legal advice on borderline cases.
\item \textbf{Governance framework.} Establish: a named senior individual for NIS compliance; documented risk management; board/executive reporting.
\item \textbf{ICO registration.} Register with the ICO as competent authority. Engage with ICO guidance on expected technical and organisational measures.
\item \textbf{Baseline security assessment.} Conduct first CAF assessment using the gap analysis in Appendix~\ref{app:gapanalysis}.
\item \textbf{Incident response capability.} Develop a plan satisfying 24/72-hour reporting. Establish CA and NCSC contacts. Create customer notification templates.
\end{enumerate}

\subsubsection{Phase 2: Medium-Term (6--18 months)}
\begin{enumerate}[label=\textbf{MSP-\arabic*.}, leftmargin=*, start=6, itemsep=2pt]
\item \textbf{Technical controls.} Implement: MFA for all admin access; client environment segmentation; encryption in transit/at rest; vulnerability management; logging with defined retention.
\item \textbf{DCS preparedness.} Where supplying OES, prepare for potential designation. Ensure contracts support regulatory cooperation and incident sharing.
\item \textbf{Customer notification framework.} Implement classification criteria, templates, escalation procedures, and evidence retention.
\item \textbf{BC/DR.} Develop and test plans addressing simultaneous multi-client disruption.
\item \textbf{Staff training.} Mandatory programme with role-specific modules for admin access holders and incident responders.
\end{enumerate}

\subsubsection{Phase 3: Steady State (18+ months)}
\begin{enumerate}[label=\textbf{MSP-\arabic*.}, leftmargin=*, start=11, itemsep=2pt]
\item \textbf{Compliance assurance.} Regular internal audit, external assessment, and continuous monitoring.
\item \textbf{Regulatory reporting.} Ongoing ICO returns and ad hoc information requests.
\item \textbf{Contract review.} Update client contracts for NIS obligations: security responsibilities, incident cooperation, audit rights, termination provisions.
\end{enumerate}

\subsection{Designated Critical Suppliers (DCSs)}

\subsubsection{Pre-Designation Preparedness}
\begin{enumerate}[label=\textbf{DCS-\arabic*.}, leftmargin=*, itemsep=2pt]
\item \textbf{Self-assess designation likelihood.} Key indicators: direct supply to known OES; NIS-dependent service delivery; absence of alternatives; high business criticality.
\item \textbf{Proactive engagement.} Engage with OES clients and CAs. Early engagement enables collaborative risk assessment.
\item \textbf{Baseline compliance.} Implement Cyber Essentials and CAF Basic Profile controls regardless of designation status.
\item \textbf{Contractual preparedness.} Strengthen agreements with OES clients: incident notification, audit access, security reporting, continuity obligations.
\end{enumerate}

\subsubsection{Post-Designation}
\begin{enumerate}[label=\textbf{DCS-\arabic*.}, leftmargin=*, start=5, itemsep=2pt]
\item \textbf{Full CAF assessment.} Against the CA-determined profile. Develop remediation plan.
\item \textbf{Reporting alignment.} Implement 24/72-hour procedures adapted to the designating CA's requirements.
\item \textbf{Ongoing assurance.} Maintain compliance, cooperate with regulatory assessments, and report regularly.
\end{enumerate}

\subsection{Implementation Timeline}

\begin{table*}[!t]
\centering
\caption{Consolidated Implementation Timeline by Entity Type}
\label{tab:timeline}
\begin{tabular}{L{1.8cm}L{4.2cm}L{4.2cm}L{4.2cm}}
\toprule
\textbf{Timeframe} & \textbf{OES} & \textbf{RMSP} & \textbf{DCS} \\
\midrule
0--3 months & Board briefing; gap analysis; supply chain inventory & Scope determination; ICO registration; governance framework & Self-assess designation likelihood; proactive engagement \\
\addlinespace
3--6 months & Incident process re-engineering; CA engagement & Baseline CAF assessment; incident response plan & Contractual preparedness; baseline security \\
\addlinespace
6--12 months & Supply chain risk assessment; DCS engagement; technical uplift & Technical controls; staff training; customer notification & (If designated) Full CAF assessment; remediation planning \\
\addlinespace
12--18 months & Code of Practice alignment; incident rehearsals & BC/DR testing; supply chain obligations; contract review & Reporting alignment; ongoing assurance cycle \\
\addlinespace
18+ months & Steady-state CAF cycle; continuous improvement & Ongoing assurance; regulatory reporting & Ongoing compliance; CA cooperation \\
\bottomrule
\end{tabular}
\end{table*}

\section{Case Studies: UK Cyber Incidents Mapped to Bill Provisions}
\label{app:casestudies}

\subsection{Case Study 1: Advanced Computer Software / NHS (August 2023)}

\textbf{What happened:} Advanced, an MSP supplying clinical software to the NHS, suffered a ransomware attack that disrupted NHS 111 services and forced multiple trusts to manual processes. Patient data was compromised. The ICO subsequently fined Advanced \pounds3.07 million under the UK GDPR \cite{cliffordchance2025}.

\textbf{Under the CS\&R Bill:} Advanced would be classified as an RMSP under ICO oversight. The following provisions would apply:

\begin{itemize}[itemsep=1pt]
\item \textit{Reporting:} 24-hour initial notification to ICO; 72-hour full report; concurrent NCSC notification; customer (NHS trust) notification. Under the existing regime, there was no NIS reporting obligation.
\item \textit{Penalties:} Up to \pounds17m / 4\% turnover---substantially exceeding the \pounds3.07m GDPR fine actually imposed.
\item \textit{DCS designation:} Advanced could have been pre-emptively designated as a critical supplier to the NHS OES, enabling regulatory oversight \textit{before} the incident.
\end{itemize}

\textbf{Key lesson:} The Bill's most important structural improvement is bringing MSPs within scope. A provider whose compromise directly disrupted a nationally significant healthcare service previously faced no NIS consequences.

\subsection{Case Study 2: Capita (March 2023)}

\textbf{What happened:} Capita, a major outsourcer to local government, the NHS, and the MoD, suffered a Black Basta ransomware attack disrupting services across multiple clients. Personal data was exfiltrated. The ICO imposed a \pounds14 million fine \cite{cliffordchance2025}.

\textbf{Under the CS\&R Bill:} Capita would be an RMSP with likely DCS designations from multiple CAs given its cross-sector client base. Multi-regulator obligations would apply simultaneously. The Secretary of State's national security powers could be triggered given MoD service disruption.

\textbf{Key lesson:} Capita illustrates \textbf{concentration risk}: a single provider failure cascading across multiple CNI sectors. The DCS mechanism and formal supply chain risk assessment provide tools for identifying and mitigating such systemic risks proactively.

\subsection{Case Study 3: Marks \& Spencer (2025)}

\textbf{What happened:} M\&S suffered a significant attack causing substantial operational disruption to retail and online operations.

\textbf{Under the CS\&R Bill:} M\&S operates in retail, which is \textit{not within scope}. The Bill imposes no obligations. However, the Secretary of State's scope expansion powers could be used to designate retail as an essential activity in future.

\textbf{Key lesson:} The strongest argument for accelerated scope expansion. Nationally significant economic disruption is not confined to currently regulated sectors.

\subsection{Case Study 4: Jaguar Land Rover (2025)}

\textbf{What happened:} A cyber attack estimated to have cost the UK economy \pounds1.9 billion disrupted manufacturing operations and supply chain coordination \cite{traverssmith2025bill}.

\textbf{Under the CS\&R Bill:} Automotive manufacturing is not in scope. However, JLR's critical IT suppliers may themselves be RMSPs or DCSs, creating indirect regulatory coverage of parts of JLR's supply chain.

\textbf{Key lesson:} When there is no regulatory penalty for poor security, the economic incentive to invest is weaker. The Bill corrects this for in-scope entities, but the JLR case demonstrates the macroeconomic cost of gaps in coverage.

\subsection{Summary}

\begin{table*}[!t]
\centering
\caption{Case Study Mapping to CS\&R Bill Provisions}
\label{tab:casesummary}
\begin{tabular}{L{1.8cm}L{1.3cm}L{1.3cm}L{2.2cm}L{2.2cm}L{2.0cm}L{2.0cm}}
\toprule
\textbf{Case} & \textbf{In Scope?} & \textbf{Entity Type} & \textbf{24/72h Reporting} & \textbf{Penalty Exposure} & \textbf{DCS Relevant?} & \textbf{SoS Powers?} \\
\midrule
Advanced & Yes & RMSP & Yes + customer notification & \pounds17m / 4\% & Yes (health CA) & Potentially \\
\addlinespace
Capita & Yes & RMSP/DCS & Yes, multi-sector & \pounds17m / 4\% per CA & Yes, multiple CAs & Yes (MoD) \\
\addlinespace
M\&S & No & N/A & None & None & Suppliers may be & Only via expansion \\
\addlinespace
JLR & No & N/A & None & None & Suppliers may be & Only via expansion \\
\bottomrule
\end{tabular}
\end{table*}

\section{Sector-Specific Action Plans}
\label{app:sectoral}

\subsection{Financial Services}
\begin{enumerate}[label=\textbf{FS-\arabic*.}, leftmargin=*, itemsep=2pt]
\item \textbf{Unified control mapping.} Map CS\&R Bill requirements (CAF v4.0) against FCA/PRA operational resilience controls and DORA. Identify shared controls and bespoke gaps.
\item \textbf{Third-party risk integration.} Combine DCS framework, DORA's TPSP register, and the Bank of England's critical third-party regime into a single programme.
\item \textbf{TLPT alignment.} Extend DORA threat-led penetration testing scope to cover CS\&R Bill in-scope systems. Use results to evidence CAF Objective B and C outcomes.
\item \textbf{Board governance.} Adopt DORA's management body accountability standard across UK operations, not merely EU-facing activities.
\item \textbf{Incident reporting coordination.} Build a unified triage function for: CS\&R 24/72h to CA + NCSC; DORA major incident to ESA; FCA/PRA SUP 15.3 notifications; customer notifications under both regimes.
\item \textbf{Concentration risk analysis.} Formally assess IT service provider dependencies and develop diversification strategies for single-points-of-failure.
\end{enumerate}

\subsection{Energy}
\begin{enumerate}[label=\textbf{EN-\arabic*.}, leftmargin=*, itemsep=2pt]
\item \textbf{OT security assessment.} Comprehensive assessment mapped to CAF v4.0 with Ofgem sector-specific guidance. Prioritise Objective B outcomes in OT environments.
\item \textbf{IT/OT segmentation.} Strengthen separation between IT and OT using ZTA micro-segmentation.
\item \textbf{Large load controller readiness.} Newly in-scope entities should achieve Cyber Essentials and CAF Basic Profile as immediate baseline.
\item \textbf{Supply chain focus.} Energy OT vendors are highly specialised with limited alternatives. Map dependencies; engage proactively on DCS designation.
\item \textbf{National security preparedness.} Establish rapid response capabilities and communication channels with Ofgem and NCSC for potential SoS directions.
\end{enumerate}

\subsection{Health}
\begin{enumerate}[label=\textbf{HE-\arabic*.}, leftmargin=*, itemsep=2pt]
\item \textbf{MSP dependency mapping.} NHS trusts should map all MSP dependencies and assess provider RMSP compliance readiness.
\item \textbf{Clinical system resilience.} Prioritise patient-facing systems. Include safe clinical downgrade procedures in incident response plans.
\item \textbf{Medical device security.} Inventory connected devices; assess against CAF Objective B; engage manufacturers on vulnerability management.
\item \textbf{Cross-trust coordination.} Strengthen shared threat intelligence, joint incident response, and coordinated DHSC/NCSC engagement.
\item \textbf{Patient data alignment.} Integrate CS\&R notification duties with Caldicott Guardian and GDPR breach notification workflows.
\end{enumerate}

\subsection{Managed Service Providers}
\begin{enumerate}[label=\textbf{MP-\arabic*.}, leftmargin=*, itemsep=2pt]
\item \textbf{Cyber Essentials as foundation.} Achieve CE and CE+ certification as an immediate baseline.
\item \textbf{Client segmentation.} Identify which clients are OES; plan for multi-regulator DCS engagement.
\item \textbf{Evidence management.} Implement structured evidence capture: control implementation, incident records, training, supply chain assessments, change management.
\item \textbf{Insurance review.} Review PI and cyber coverage against the \pounds17m/4\% penalty framework. Ensure policy conditions are compatible with 24/72h timelines.
\item \textbf{Client communication.} Position NIS compliance as a competitive differentiator and client assurance.
\item \textbf{M\&A considerations.} Incorporate NIS compliance into acquisition due diligence. Non-compliant targets represent regulatory risk.
\end{enumerate}

\section{Gap Analysis and Self-Assessment Tool}
\label{app:gapanalysis}

Score each item as \textbf{A} (Achieved), \textbf{PA} (Partially Achieved), or \textbf{NA} (Not Achieved). Priority: \textbf{H} (High---fundamental gap; regulatory risk), \textbf{M} (Medium---partial; improvement needed), \textbf{L} (Low---minor enhancement).

\subsection{Objective A: Managing Security Risk}

\begin{table*}[!t]
\centering
\caption{Gap Analysis: Objective A --- Managing Security Risk}
\label{tab:gapa}
\begin{tabular}{L{0.4cm}L{3.0cm}L{6.0cm}L{0.8cm}L{2.2cm}}
\toprule
\textbf{\#} & \textbf{Area} & \textbf{Expected Standard} & \textbf{Score} & \textbf{Priority} \\
\midrule
A1 & Governance & Named senior individual for NIS compliance; board reporting; defined risk appetite & & \\
\addlinespace
A2 & Risk management & Documented, threat-informed risk assessment covering all NIS systems; regular review & & \\
\addlinespace
A3 & Asset management & Complete NIS asset inventory; criticality classification; configuration management & & \\
\addlinespace
A4 & Supply chain risk & Supplier register; dependency and substitutability assessment; DCS preparedness & & \\
\addlinespace
A5 & Information sharing & NCSC early warning registration; sector ISAC membership; threat intelligence consumption & & \\
\bottomrule
\end{tabular}
\end{table*}

\subsection{Objective B: Protecting Against Cyber Attack}

\begin{table*}[!t]
\centering
\caption{Gap Analysis: Objective B --- Protecting Against Cyber Attack}
\label{tab:gapb}
\begin{tabular}{L{0.4cm}L{3.0cm}L{6.0cm}L{0.8cm}L{2.2cm}}
\toprule
\textbf{\#} & \textbf{Area} & \textbf{Expected Standard} & \textbf{Score} & \textbf{Priority} \\
\midrule
B1 & Identity \& access & MFA for all admin/remote access; least privilege; PAM; dynamic access decisions & & \\
\addlinespace
B2 & Data security & Encryption at rest and in transit; data classification; DLP controls & & \\
\addlinespace
B3 & System security & Secure configuration baselines; patching; application security; hardening & & \\
\addlinespace
B4 & Network security & Micro-segmentation; ZTA principles; ingress/egress controls; supply chain isolation & & \\
\addlinespace
B5 & Resilient systems & Redundancy; failover; tested DR; defined RTOs/RPOs & & \\
\addlinespace
B6 & Staff training & Role-appropriate security training; phishing awareness; reporting culture & & \\
\bottomrule
\end{tabular}
\end{table*}

\subsection{Objective C: Detecting Cyber Security Events}

\begin{table*}[!t]
\centering
\caption{Gap Analysis: Objective C --- Detecting Events}
\label{tab:gapc}
\begin{tabular}{L{0.4cm}L{3.0cm}L{6.0cm}L{0.8cm}L{2.2cm}}
\toprule
\textbf{\#} & \textbf{Area} & \textbf{Expected Standard} & \textbf{Score} & \textbf{Priority} \\
\midrule
C1 & Security monitoring & Continuous monitoring; SIEM or equivalent; log retention; real-time alerting & & \\
\addlinespace
C2 & Anomaly detection & Behavioural baselines; automated deviation alerting; threat hunting capability & & \\
\addlinespace
C3 & Threat intelligence & Operationalised threat feeds; NCSC advisories; sector-specific intelligence & & \\
\bottomrule
\end{tabular}
\end{table*}

\subsection{Objective D: Minimising Impact of Incidents}

\begin{table*}[!t]
\centering
\caption{Gap Analysis: Objective D --- Minimising Impact}
\label{tab:gapd}
\begin{tabular}{L{0.4cm}L{3.0cm}L{6.0cm}L{0.8cm}L{2.2cm}}
\toprule
\textbf{\#} & \textbf{Area} & \textbf{Expected Standard} & \textbf{Score} & \textbf{Priority} \\
\midrule
D1 & Incident response & Documented IRP; defined roles; tested within 12 months; 24/72h capability & & \\
\addlinespace
D2 & Notification & Concurrent CA + NCSC within 24h; customer notification; SoS escalation path & & \\
\addlinespace
D3 & Recovery & Tested recovery procedures; backup integrity verification; defined RTOs/RPOs & & \\
\addlinespace
D4 & Lessons learned & Post-incident review; remediation tracking; anonymised sharing within sector & & \\
\bottomrule
\end{tabular}
\end{table*}

\subsection{Bill-Specific Obligations}

\begin{table*}[!t]
\centering
\caption{Gap Analysis: CS\&R Bill-Specific Requirements}
\label{tab:gapbill}
\begin{tabular}{L{0.4cm}L{3.0cm}L{6.0cm}L{0.8cm}L{2.2cm}}
\toprule
\textbf{\#} & \textbf{Area} & \textbf{Expected Standard} & \textbf{Score} & \textbf{Priority} \\
\midrule
E1 & Registration & Registered with relevant CA; fee arrangements in place & & \\
\addlinespace
E2 & Codes of Practice & Aware of applicable Codes; mapped to controls; evidenced compliance & & \\
\addlinespace
E3 & SoS direction readiness & Awareness of emergency direction powers; rapid response capability & & \\
\addlinespace
E4 & Cross-border compliance & Dual DORA/CS\&R framework (if applicable); information sharing agreements & & \\
\addlinespace
E5 & Cost recovery & Budget for regulatory fees; understanding of CA fee methodology & & \\
\bottomrule
\end{tabular}
\end{table*}

\subsection{Using This Assessment}

Convene a cross-functional team including: the CISO or equivalent; head of IT operations; head of legal/compliance; a risk function representative; and where applicable, supply chain management. Assess each NIS-dependent system individually, not merely at organisational level.

The completed assessment provides: a structured evidence base for regulatory engagement; a prioritised remediation roadmap; a baseline for measuring progress; and board-level reporting material demonstrating due diligence.

\begin{actionbox}
This gap analysis should be completed as an immediate priority. Even if the Bill has not yet received Royal Assent, the assessment provides value as a structured review of your organisation's cyber resilience posture against the national benchmark.
\end{actionbox}

\end{document}